\documentclass[aps,prfluids,twocolumn,superscriptaddress,longbibliography]{revtex4-1}

\usepackage{graphicx}
\usepackage{MnSymbol}
\usepackage{amsmath}
\usepackage{amsfonts}
\usepackage{natbib}
\usepackage{soul}
\usepackage{color}
\usepackage{subcaption}
\usepackage{placeins}
\usepackage{mathtools,tikz,caption}
\usepackage{bm}
\usepackage{epstopdf}

\DeclareRobustCommand\sampleline[1]{%
  \tikz\draw[#1] (0,0) (0,\the\dimexpr\fontdimen22\textfont2\relax)
  -- (1em,\the\dimexpr\fontdimen22\textfont2\relax);%
}

\begin{document}

\preprint{}

\title{Direct assessment of Kolmogorov's first refined similarity hypothesis}


\author{John M. Lawson}
\email{john.lawson@ds.mpg.de}
\affiliation{Max Planck Institute for Dynamics and Self-Organisation}
\affiliation{University of Southampton}

\author{Eberhard Bodenschatz}
\affiliation{Max Planck Institute for Dynamics and Self-Organisation}

\author{Anna N. Knutsen}
\affiliation{Norwegian University of Science and Technology}

\author{James R. Dawson}
\affiliation{Norwegian University of Science and Technology}

\author{Nicholas A. Worth}
\affiliation{Norwegian University of Science and Technology}



\newcommand{\eav}[1]{\langle #1 \rangle}
\newcommand{\tav}[1]{\overline{#1}}
\newcommand{\sav}[1]{\langle #1 \rangle_{\Omega}}
\newcommand{\vav}[1]{{#1}_{r}}
\newcommand{\bhat}[1]{ \hat{\bm{#1}} }

\newcommand{\cA}[1]{{\color[rgb]{0.00, 0.00, 0.57}#1}}
\newcommand{\cB}[1]{{\color[rgb]{0.00, 0.00, 0.85}#1}}
\newcommand{\cC}[1]{{\color[rgb]{0.00, 0.28, 0.85}#1}}
\newcommand{\cD}[1]{{\color[rgb]{0.00, 0.57, 0.85}#1}}
\newcommand{\cE}[1]{{\color[rgb]{0.00, 0.85, 0.85}#1}}
\newcommand{\cF}[1]{{\color[rgb]{0.28, 0.85, 0.57}#1}}
\newcommand{\cG}[1]{{\color[rgb]{0.57, 0.85, 0.28}#1}}
\newcommand{\cH}[1]{{\color[rgb]{0.85, 0.85, 0.00}#1}}
\newcommand{\cI}[1]{{\color[rgb]{0.85, 0.57, 0.00}#1}}
\newcommand{\cJ}[1]{{\color[rgb]{0.85, 0.28, 0.00}#1}}

\date{\today}

\begin{abstract}

Using volumetric velocity data from a turbulent laboratory water flow and numerical simulations of homogeneous, isotropic turbulence, we present a direct experimental and numerical assessment of Kolmogorov's first refined similarity hypothesis based on three-dimensional measurements of the local energy dissipation rate $\epsilon_r$ measured at dissipative scales $r$.
We focus on the properties of the stochastic variables $V_L = \Delta u(r)/(r \epsilon_r)^{1/3}$ and $V_T = \Delta v(r)/(r\epsilon_r)^{1/3}$, where $\Delta u(r)$ and $\Delta v(r)$ are longitudinal and transverse velocity increments.
Over one order of magnitude of scales $r$ within the dissipative range, the distributions of $V_L$ and $V_T$ from both experiment and simulation collapse when parameterised by a suitably defined local Reynolds number, providing the first conclusive experimental evidence in support of the first refined similarity hypothesis and its universality.

\end{abstract}


\maketitle


Obtaining a universal statistical description of hydrodynamic turbulence has been a widely-pursued yet elusive objective within fluid mechanics.
Kolmogorov's refined similarity hypotheses represent one such seminal attempt \citep{Kolmogorov1962}, which underpins the modern understanding of intermittency in small scale turbulence \citep{SreenivasanAntonia1997}.
This phenomenon directly influences, amongst others, the efficiency of rain formation in clouds \citep{Shaw2003}, the production of pollutants in combustion processes \citep{Sreenivasan2004} and the propagation of sound and light through the atmosphere \citep{Tatarskii1971,Wilson1996}.
In this Rapid Communication, we overcome previous technical limitations to provide a quantitative and direct experimental assessment of the validity of the first refined similarity hypothesis with back-to-back comparisons against numerical simulations to examine their universality.

The similarity hypotheses describe turbulent flows in terms of velocity differences, or increments, $\Delta \bm{u} = \bm{u}(\bm{x},t) - \bm{u}(\bm{x}',t)$ between simultaneously measured pairs of points in the flow, where the spatial separation $\bm{r} = \bm{x}-\bm{x}'$ is much smaller than the energy injection scale $L$.
In their simplest formulation, known as K41 \citep{Kolmogorov1941a}, the distribution of velocity increments is prescribed by the scale $r=|\bm{r}|$, the average rate of kinetic energy dissipation $\langle \epsilon \rangle$ and the fluid kinematic viscosity $\nu$.
Laboratory and numerical experiments now widely confirm departures from the K41 scaling  \citep{Anselmet1984,Nelkin1994,SreenivasanAntonia1997}.
The essence of this deviation was first articulated by Landau \citep{Landau1987}, who remarked that whilst the increment distribution may plausibly uniquely depend upon a temporally localised average of the energy dissipation rate $\epsilon(\bm{x},t) = \nu \Sigma_{i,j}(\partial u_i/\partial x_j + \partial u_j/\partial x_i)^2/2$, the distribution law of $\Delta\bm{u}$ must depend upon the fluctuation of this local average over time, which may in turn depend upon the whims and fancies of the largest scale motions that feed the turbulence its energy.

Landau's criticisms are accounted for in the refined similarity scaling \citep{Kolmogorov1962,Oboukhov1962}, known as K62, by substituting $\eav{\epsilon}$ for a local dissipation rate $\epsilon_r$ 
\begin{equation}
\label{eqn:local-dissipation}
\epsilon_r(\bm{X},r,t) = 
\frac{6}{\pi r^3} \iiint_{|\bm{y}| \le r/2} \mathrm{d}\bm{y}~\epsilon(\bm{X}+\bm{y},t)
\end{equation} 
which is a spatial average of the instantaneous energy dissipation field over a sphere whose poles are defined by $\bm{x}$ and $\bm{x}'$, centered at  $\bm{X}=(\bm{x}+\bm{x}')/2$ with diameter $r$ \footnote{This definition is consistent with Oboukhov's formulation; in Kolmogorov's formulation, the averaging volume is a sphere of radius $r$ centred at $\bm{x}$.}.
This permits a characteristic velocity scale $U_{r} \equiv (r\vav{\epsilon})^{1/3}$ to be constructed local to the position, scale and time defined by $(\bm{X},r,t)$.
The two postulates of refined similarity \cite{Kolmogorov1962}, known as K62, can then be formulated as follows for some randomly oriented $\bm{r}$ such that $r \ll L$ \citep{Stolovitzky1992,Iyer2017}: (i) the distribution of $\bm{V} = \Delta\bm{u}/U_r$ depends only upon the local Reynolds number $Re_r = U_r r/\nu$ and (ii) is independent of $Re_r$ when $Re_r \gg 1$.

Fifty-six years hence, the experimental evidence for K62 is far from conclusive and has focused exclusively on the second postulate applied to a single component of $\bm{V}$ parallel to $\bm{r}$ \citep{Stolovitzky1992,Thoroddsen1992,Praskovsky1992,Thoroddsen1995,Qian1996}.
Early reports \citep{Stolovitzky1992,Thoroddsen1992,Praskovsky1992} offered tentative support for the second postulate. However, closer inspection has revealed that the available experimental data are inconsistent with the implications of combining the second K62 postulate with three plausible models for the distribution of $\vav{\epsilon}$ \citep{Qian1996}.
The discrepancy lies in the use of two simplifications used to obtain $\vav{\epsilon}$ experimentally, wherein volume averaging is replaced by one-dimensional (1D) line averaging and a 1D surrogate $\epsilon' = 15\nu(\partial u_1/\partial x_1)^2$ is substituted for $\epsilon$.
The use of the surrogate $\epsilon'$ severely distorts the available experimental evidence, since its use weakens the dependence between $\Delta \bm{u}$ and $(r\epsilon_r)^{1/3}$ \citep{Wang1996} and the dependence all but disappears when other, plausible surrogates for $\epsilon$ are used \citep{Thoroddsen1995,Chen1995}. 
One or both of these simplifications have also been employed in numerical studies on the K62 postulates \citep{Chen1993,Chen1995,Wang1996,Schumacher2007}. 
Two notable exceptions are Refs. \citep{Iyer2015} and \citep{Iyer2017}.
These have provided the first evidence for K62 scaling obtained by direct numerical simulation (DNS) of the Navier-Stokes equations using 3D averages and argue that previous numerical evidence disfavouring the K62 postulates \citep{Schumacher2007} stems from the inappropriate use of 1D averaging.
The question therefore arises whether the same distribution of $\bm{V}$ found in numerical experiments can also be found in nature, which invariably lacks the statistical symmetries of such simulations that may influence both the distribution of $\bm{V}$ and its scaling \citep{Iyer2017}.

In the following, we address the deficiencies of previous experiments by directly examining the first K62 postulate without resort to surrogates.
This is achieved using a recently developed technique \citep{Lawson2014} to make volumetric velocity measurements capable of directly measuring $\vav{\epsilon}$ in a volume large enough to test the first K62 posulate across a decade of scales.
We complement this data with back-to-back comparisons against direct numerical simulations of homogeneous, isotropic turbulence \citep{Cardesa2017} to test the universality of the statistics of $\bm{V}$.

We measured the turbulence in a $1\textrm{cm}^3$ measurement volume near the mean-field stagnation point of a von-K{\'a}rm{\'a}n swirling water flow \citep{Xu2007, Xu2011} using Scanning Particle Image Velocimetry (PIV) \citep{Lawson2014}. 
This volume is small in comparison to the characteristic size of the energy containing motions $L =  u'^3/ \langle\epsilon\rangle \approx 77\textrm{mm}$, where $u'^2 = \langle u_i' u_i' \rangle/3$ is the mean-square velocity fluctuation.
The Taylor microscale Reynolds number was $R_\lambda \approx 200$.
The working fluid, deionised water, was seeded with $6\mu\textrm{m}$ diameter PMMA microspheres with specific gravity 1.22, which are $35$ times smaller than the Kolmogorov lengthscale $\eta = (\nu^3/\eav{\epsilon})^{1/4} \approx 210\mu\textrm{m}$ and act as passive flow tracers.
The flow was illuminated with a $4.7\eta$ thick laser light sheet from a $90\textrm{W}$, pulsed, Nd:YAG laser, which was rapidly scanned across the measurement volume 250 times per second using a galvanometer mirror scanner.
A pair of Phantom v640 high-speed cameras recorded the forward-scattered light at $\pm 45^\circ$ to the sheet at $15\textrm{kHz}$ with a resolution of $512 \times 512$ pixels.
Each was equipped with $200\textrm{mm}$ focal length macro lenses and $2\times$ teleconverters, providing $1:2$ optical magnification and a spatial resolution of $20\mu\textrm{m}$ per pixel. 
For each sample, we stored five scans with 54 consecutive images each, corresponding to a spacing between parallel laser sheets of $1.3\eta$. 

The distribution of tracers was tomographically reconstructed in a discretised volume of $521 \times 513 \times 515$ voxels using the method described in Ref. \citep{Lawson2014}.
The scanning method enabled us to make reconstructions with a high seeding concentration of around 1 particle per $(1.4\eta)^3$.
Reconstructions from sequential scans were cross-correlated with a multi-pass PIV algorithm described in Ref. \citep{Lawson2014} with an interrogation window size of $3.2\eta$ and corrections applied to account for the finite acquisition time.
This yielded volumetric measurements of the velocity field in a $(42\eta)^3$ volume on a regular grid with spacing $0.8\eta$, from which we obtained the full velocity gradient tensor and hence the dissipation field using a least-squares finite difference stencil \citep{Raffel2007}.
We gathered samples at $4.5$ second intervals during the continuous operation of the experimental facility for $11$ days. The water temperature was maintained at $21.2 \pm 0.5^\circ \textrm{C}$ by a heat exchanger, seeding concentration was maintained at $24$ hour intervals and scanning PIV calibration accuracy was maintained using the method in Ref. \citep{Knutsen2017}.
This yielded $2\times 10^5$ statistically independent volumetric snapshots of the velocity and dissipation fields.

We complement our experimental dataset with statistics obtained from publicly available DNS of forced, homogeneous isotropic turbulence lasting $66$ large eddy turnover times \citep{Cardesa2017}.
The pseudospectral simulation solved the incompressible, Navier-Stokes equations on a grid of $1024^3$ collocation points in a triply-periodic domain with a fixed energy injection rate forcing and maximum resolvable wavenumber $k_{max}\eta = 2$.
Whilst the Taylor microscale Reynolds number $R_\lambda \approx 315$ has been surpassed by other works, the long duration of this simulation allowed us to gather well converged statistics.
Velocity gradients were evaluated spectrally to obtain $\epsilon$.
The local dissipation rate $\epsilon_{r}$ (\ref{eqn:local-dissipation}) was obtained using the spectral method in Ref. \citep{BorueOrszag1996}.
Following \citep{Iyer2015}, triplets of longitudinal $\Delta u(\bm{X},\bm{r},t) = \Delta\bm{u}\cdot\bhat{r}$ and transverse $\Delta v(\bm{X},\bm{r},t) = \Delta\bm{u}\cdot(\bm{e}_j\times \bhat{r})$ velocity increments were evaluated for $\bhat{r}$ oriented in each of the three principal grid directions $\bm{e}_i$ ($i\ne j$) over separations $r/\eta$ of $3.0, 5.9, 8.9, 11.8, 17.8, 23.7, 32.6$, corresponding to logarithmically spaced, even multiples of the grid spacing.
Statistics were evaluated for each grid point in $66$ snapshots of the flow field spaced evenly in time over the simulated time interval.

In contrast to the numerical simulation data, the data from our von-K{\'a}rm{\'a}n mixing tank exhibit a statistical axisymmetry aligned with the axis of the counter-rotating disks \citep{Voth1998,Voth2002,Lawson2015}.
We therefore adopt a careful definition of our statistical ensemble of $\Delta u(\bm{X},\bm{r},t)$, $\Delta v(\bm{X},\bm{r},t)$ and $\vav{\epsilon}(\bm{X},r,t)$ in order to recover the isotropic scaling behaviour.
For a single point $\bm{X}$ near the mean-flow stagnation point, we evaluate the longitudinal and transverse velocity increments over $2940$ orientations of the separation vector $\bm{r}$ uniformly spaced over the surface of a sphere of diameter $r$. 
Numerically, this is achieved using a cubic spline interpolation of the velocity and dissipation field at scales $r/\eta$ chosen from the geometric series $1.5, 2.1, ... 36.2$. 
Statistics are then gathered over each of the $2\times 10^5$ realisations of the flow.
This angle-averaging of statistics is directly related to the SO(3) decomposition \citep{Arad1998,Biferale2005}, which enables the recovery of isotropic scaling properties in flows with statistical anisotropies \citep{Arad1999,Kurien2000}.

\begin{figure}
\begin{subfigure}{0.5\textwidth}
	\flushleft
	\includegraphics[trim=0cm 0cm 0cm 0cm, clip, width=0.98\textwidth]{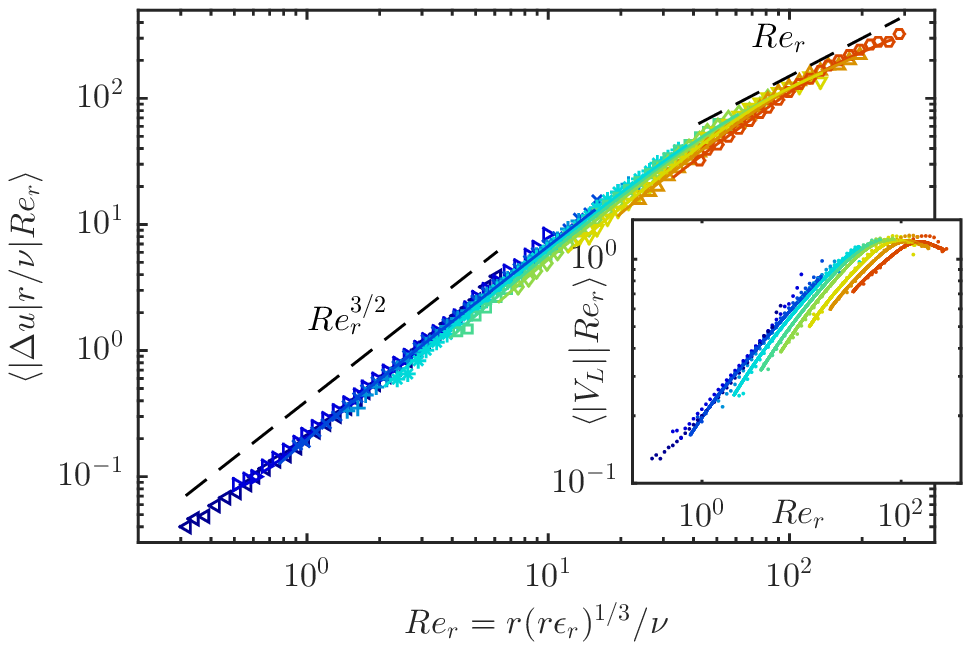}
	\caption{Longitudinal velocity increment}
	\label{fig:DLr-cond-Rer}
\end{subfigure}
\begin{subfigure}{0.5\textwidth}
	\flushleft
	\includegraphics[trim=0cm 0cm 0cm 0cm, clip, width=0.98\textwidth]{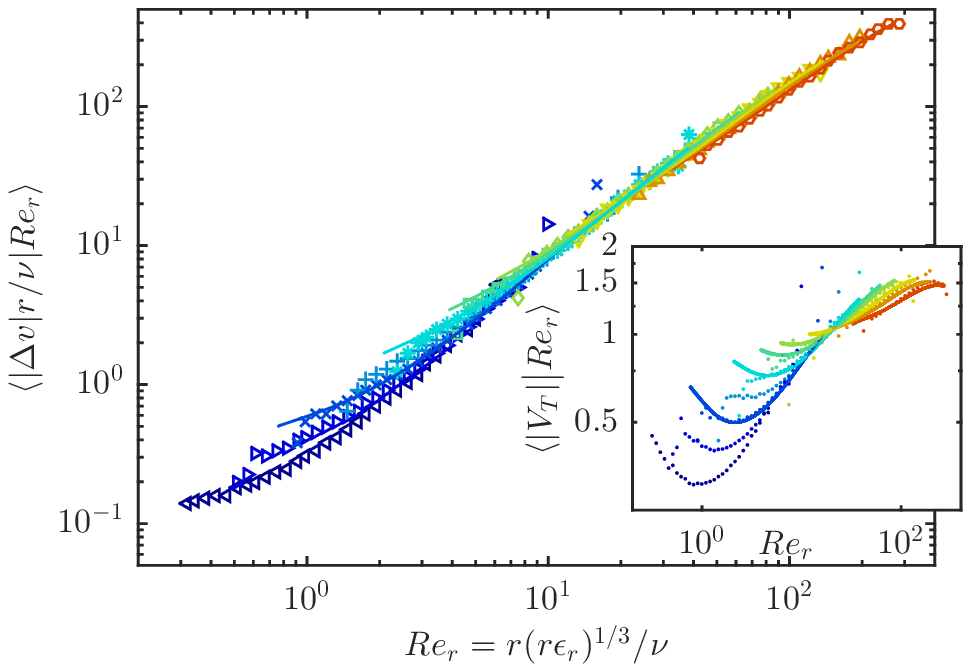} 
	\caption{Transverse velocity increment}
	\label{fig:DNr-cond-Rer}
\end{subfigure}
\caption{Scaling of (\subref{fig:DLr-cond-Rer}) longitudinal and (\subref{fig:DNr-cond-Rer}) transverse velocity increment magnitude, for fixed scale $r/\eta$, given the local Reynolds number $Re_r$. 
Symbols $\cA\medtriangleleft$, $\cB\medtriangleright$, $\cC\times$, $\cD+$, $\cE\ast$, $\cF\medsquare$, $\cG\meddiamond$, $\cH\medtriangledown$, $\cI\medtriangleup$, $\cJ\medcircle$ show experimental data at ten scales $r$ logarithmically spaced between $1.5$ and $36.2\eta$.
Solid lines show data from DNS at comparable scales.
Inset: conditional average magnitudes of $\eav{|V_L|\big|Re_r}$ and $\eav{|V_T|\big|Re_r}$.}
\label{fig:DLr-DNr-cond-Rer}
\end{figure}

To test the first K62 postulate, we consider the conditional expectations of the form
\begin{align}
\label{eqn:VL}
\langle r |\Delta u| / \nu \big| Re_{r} \rangle &= \langle |V_L| \big| Re_r\rangle Re_r \\
\label{eqn:VT}
\langle r |\Delta v| / \nu \big| Re_{r} \rangle &= \langle |V_T| \big| Re_r\rangle Re_r.
\end{align}
Under the first K62 postulate, these conditional averages should only depend upon $Re_r$.

Figure \ref{fig:DLr-cond-Rer} shows the conditional average (\ref{eqn:VL}) of the magnitude of the longitudinal velocity increment given the local Reynolds number based on $U_r$.
At comparable scales $r/\eta$, the experimental and numerical data are in close, quantitative agreement.
For each curve with fixed $r$ and $\nu$, we are effectively examining the conditional expectation of $|\Delta u|$ for different local characteristic velocity scales $U_r$.
At small $Re_r$ the data are in close agreement with the scaling $|\Delta u|r \sim Re_{r}^{3/2}$, which is expected from a Taylor series expansion at small $r$ \citep{Wang1996}.
At larger $Re_r$, the scaling approaches $|\Delta u|r \sim Re_r$, which is expected from the second K62 postulate.
If the first postulate holds exactly, given that $r \ll L$, we should expect that (\ref{eqn:VL}) only depends on $Re_{r}$.
Instead, we notice that a systematic dependence upon the scale $r$ is retained, which becomes less significant as the local Reynolds number is increased.

In contrast, Figure \ref{fig:DNr-cond-Rer} shows the equivalent conditional average (\ref{eqn:VT}) for the transverse velocity increments.
Again, there is excellent agreement between numerics and experiment.
Good collapse across scale $r$ is observed for $Re_r \gtrsim 10$.
At smaller $Re_r$, the collapse across scale is less compelling.
This may be anticipated from a consideration of the limiting behaviour of $V_T$ at small $r$.
Based on a Taylor series expansion of $\Delta v$ with orientation averaging, we obtain $\langle V_T^2 | Re_r \rangle = Re_r/20 + Re_r \langle \Omega/\epsilon \big| Re_r\rangle / 12$, where $\Omega$ is the enstrophy $\Omega = \nu(\nabla \times \bm{u})^2$.
It follows that in the limit of $r \rightarrow 0$, $\langle \Omega \big| \epsilon \rangle$ must scale linearly with $\epsilon$ for $\langle V_T^2 \big| Re_r\rangle$ to depend only upon $Re_r$.
Such a linear scaling has been shown to hold in relatively active dissipative regions $\epsilon > \langle\epsilon\rangle$ of homogeneous isotropic turbulence, but breaks down for $\epsilon \ll \eav{\epsilon}$  \citep{Donzis2008}.
The discrepancy may be resolved as the Taylor microscale Reynolds number is increased \citep{Yeung2015}.

\begin{figure}[h!]
\begin{subfigure}{0.27\textwidth}
	\includegraphics[trim=0cm 0cm 0cm 0cm, clip, width=\textwidth]{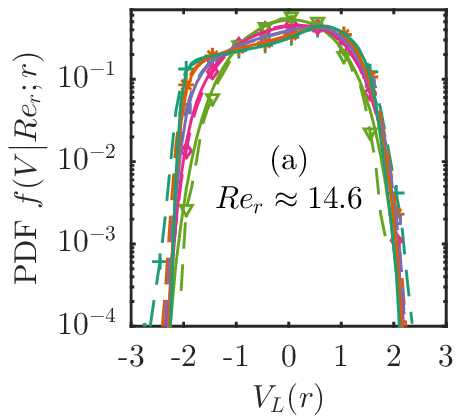}
\end{subfigure}
\begin{subfigure}{0.197\textwidth}
	\includegraphics[trim=0cm 0cm 0cm 0cm, clip, width=\textwidth]{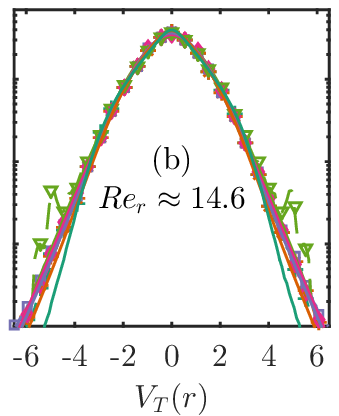}
\end{subfigure}
\begin{subfigure}{0.27\textwidth}
	\includegraphics[trim=0cm 0cm 0cm 0cm, clip, width=\textwidth]{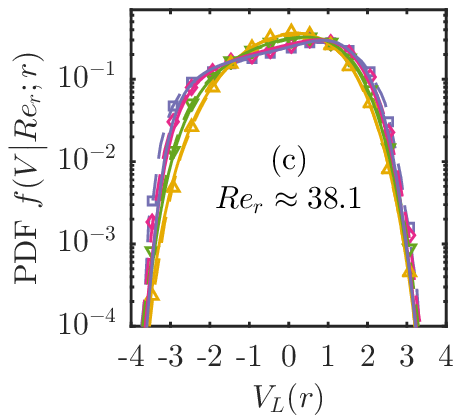}
\end{subfigure}
\begin{subfigure}{0.201\textwidth}
	\includegraphics[trim=0cm 0cm 0cm 0cm, clip, width=\textwidth]{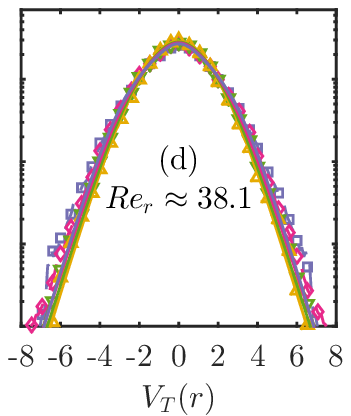}
\end{subfigure}
\begin{subfigure}{0.27\textwidth}
	\includegraphics[trim=0cm 0cm 0cm 0cm, clip, width=\textwidth]{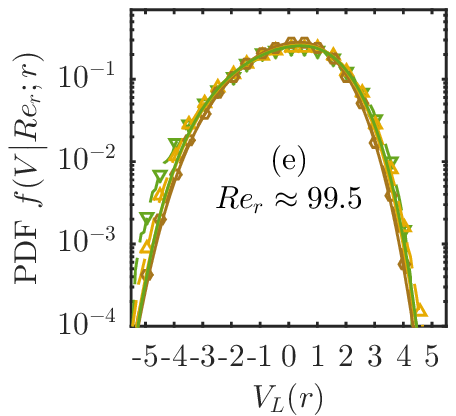}
\end{subfigure}
\begin{subfigure}{0.205\textwidth}
	\includegraphics[trim=0cm 0cm 0cm 0cm, clip, width=\textwidth]{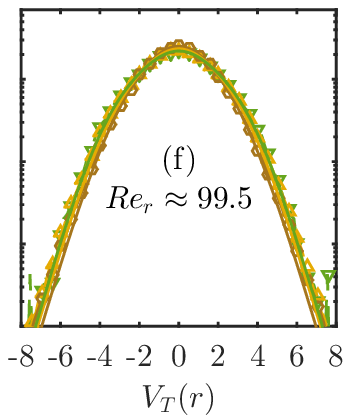}
\end{subfigure}
\caption{Conditional distribution of dimensionless longitudinal increments $V_L$ (left panels) and transverse increments $V_T$ (right panels) at different scales $r/\eta$ given fixed local Reynolds number.
In each pair of panels, the local Reynolds number, number of experimental curves, and minimum and maximum scales $r/\eta$ are respectively (a,b) $14.6, 5, 4.3, 17.8$ (c,d) $38.1, 4, 8.8, 25.4$ and (e,f) $99.5, 3, 17.8, 36.2$.
The local Reynolds number at each scale is matched to within $5\%$ of the nominal value.
Lines show: (\sampleline{dashed}) experimental (\sampleline{solid}) DNS data.
Symbols and colour denote scale $r/\eta$. Markers are as in Figure \ref{fig:DLr-DNr-cond-Rer}.}
\label{fig:V-pdf-cond-Re}
\end{figure}

As a detailed, direct test of first postulate, Figure \ref{fig:V-pdf-cond-Re} shows the conditional distribution of $V_L$ and $V_T$ given the local Reynolds number over different spatial scales $r$.
For both longitudinal and transverse increments, we observe good, quantitative agreement between experimental and numerical data at comparable scales $r/\eta$ and matched local Reynolds numbers.
We first consider the longitudinal velocity increment.
The left panels of Figure \ref{fig:V-pdf-cond-Re} demonstrate that the distribution of $V_L$ largely collapses across scale when conditioned upon the local Reynolds number.
The collapse improves as the local Reynolds number is made larger.
The scale dependence of the conditional distribution of $V_L$ appears to be stronger in our data than the numerical simulation results of \citet{Iyer2015}.
This may be due to the smaller scale separation $r/L$ achieved in the present experimental study.
We offer an additional remark that, when $U_r$ is instead based on local averages of the pseudodissipation $\phi = \nu A_{ij}A_{ij}$, an improved collapse is observed for the longitudinal velocity increment.

The right panels of Figure \ref{fig:V-pdf-cond-Re} show the equivalent conditional distribution for the transverse velocity increment.
For fixed $Re_r$, the transverse increments show an improved collapse across scale in comparison to their longitudinal counterparts.
This confirms the approximate validity of the first refined similarity hypothesis for transverse velocity increments.

The application of scanning PIV has allowed us to directly examine the first K62 postulate in a laboratory flow using three-dimensional, local averages of the dissipation, thereby resolving the surrogacy issue which has confounded previous experimental investigations.
We have complemented our experimental analysis with back-to-back comparisons against high-resolution DNS of homogeneous isotropic turbulence.
We observe that the distributions of $V_L$ and $V_T$ and their average magnitudes are in close agreement between both flows when the local Reynolds number and scale are matched.
The first postulate is shown to approximately hold for both longitudinal and transverse increments, with improved agreement found for larger local Reynolds numbers.
Our study provides the first unambiguous experimental evidence to demonstrate that a detailed, universal description of high Reynolds number turbulence may at last be within grasp.

\begin{acknowledgments}
All authors designed the research; J.L. and A.K. performed the experiments; J.L. analysed data and wrote the paper; E.B., J.D. and N.W. edited the paper.
The authors gratefully acknowledge the support of the Max Planck Society and EuHIT: European High-Performance Infrastructures in Turbulence, funded under the European Union's Seventh Framework Programme (FP7/2007-2013) Grant Agreement No. 312778. 
We thank M. Wilczek and C.C. Lalescu for their helpful comments.
\end{acknowledgments}


\begin{thebibliography}{39}%
\makeatletter
\providecommand \@ifxundefined [1]{%
 \@ifx{#1\undefined}
}%
\providecommand \@ifnum [1]{%
 \ifnum #1\expandafter \@firstoftwo
 \else \expandafter \@secondoftwo
 \fi
}%
\providecommand \@ifx [1]{%
 \ifx #1\expandafter \@firstoftwo
 \else \expandafter \@secondoftwo
 \fi
}%
\providecommand \natexlab [1]{#1}%
\providecommand \enquote  [1]{``#1''}%
\providecommand \bibnamefont  [1]{#1}%
\providecommand \bibfnamefont [1]{#1}%
\providecommand \citenamefont [1]{#1}%
\providecommand \href@noop [0]{\@secondoftwo}%
\providecommand \href [0]{\begingroup \@sanitize@url \@href}%
\providecommand \@href[1]{\@@startlink{#1}\@@href}%
\providecommand \@@href[1]{\endgroup#1\@@endlink}%
\providecommand \@sanitize@url [0]{\catcode `\\12\catcode `\$12\catcode
  `\&12\catcode `\#12\catcode `\^12\catcode `\_12\catcode `\%12\relax}%
\providecommand \@@startlink[1]{}%
\providecommand \@@endlink[0]{}%
\providecommand \url  [0]{\begingroup\@sanitize@url \@url }%
\providecommand \@url [1]{\endgroup\@href {#1}{\urlprefix }}%
\providecommand \urlprefix  [0]{URL }%
\providecommand \Eprint [0]{\href }%
\providecommand \doibase [0]{http://dx.doi.org/}%
\providecommand \selectlanguage [0]{\@gobble}%
\providecommand \bibinfo  [0]{\@secondoftwo}%
\providecommand \bibfield  [0]{\@secondoftwo}%
\providecommand \translation [1]{[#1]}%
\providecommand \BibitemOpen [0]{}%
\providecommand \bibitemStop [0]{}%
\providecommand \bibitemNoStop [0]{.\EOS\space}%
\providecommand \EOS [0]{\spacefactor3000\relax}%
\providecommand \BibitemShut  [1]{\csname bibitem#1\endcsname}%
\let\auto@bib@innerbib\@empty
\bibitem [{\citenamefont {Kolmogorov}(1962)}]{Kolmogorov1962}%
  \BibitemOpen
  \bibfield  {author} {\bibinfo {author} {\bibfnamefont {A~N}\ \bibnamefont
  {Kolmogorov}},\ }\bibfield  {title} {\enquote {\bibinfo {title} {{A
  refinement of previous hypotheses concerning the local structure of
  turbulence in a viscous incompressible fluid at high Reynolds number}},}\
  }\href@noop {} {\bibfield  {journal} {\bibinfo  {journal} {Journal of Fluid
  Mechanics}\ }\textbf {\bibinfo {volume} {13}},\ \bibinfo {pages} {82}
  (\bibinfo {year} {1962})}\BibitemShut {NoStop}%
\bibitem [{\citenamefont {Sreenivasan}\ and\ \citenamefont
  {Antonia}(1997)}]{SreenivasanAntonia1997}%
  \BibitemOpen
  \bibfield  {author} {\bibinfo {author} {\bibfnamefont {K.~R.}\ \bibnamefont
  {Sreenivasan}}\ and\ \bibinfo {author} {\bibfnamefont {R.~A.}\ \bibnamefont
  {Antonia}},\ }\bibfield  {title} {\enquote {\bibinfo {title} {{The
  phenomenology of small-scale turbulence}},}\ }\href@noop {} {\bibfield
  {journal} {\bibinfo  {journal} {Annual Review of Fluid Mechanics}\ }\textbf
  {\bibinfo {volume} {29}},\ \bibinfo {pages} {435--472} (\bibinfo {year}
  {1997})}\BibitemShut {NoStop}%
\bibitem [{\citenamefont {Shaw}(2003)}]{Shaw2003}%
  \BibitemOpen
  \bibfield  {author} {\bibinfo {author} {\bibfnamefont {Raymond~A.}\
  \bibnamefont {Shaw}},\ }\bibfield  {title} {\enquote {\bibinfo {title}
  {{Particle-Turbulence Interactions in Atmospheric Clouds}},}\ }\href@noop {}
  {\bibfield  {journal} {\bibinfo  {journal} {Annual Review of Fluid
  Mechanics}\ }\textbf {\bibinfo {volume} {35}},\ \bibinfo {pages} {183--227}
  (\bibinfo {year} {2003})}\BibitemShut {NoStop}%
\bibitem [{\citenamefont {Sreenivasan}(2004)}]{Sreenivasan2004}%
  \BibitemOpen
  \bibfield  {author} {\bibinfo {author} {\bibfnamefont {K.~R.}\ \bibnamefont
  {Sreenivasan}},\ }\bibfield  {title} {\enquote {\bibinfo {title} {{Possible
  effects of small-scale intermittency in turbulent reacting flows}},}\
  }\href@noop {} {\bibfield  {journal} {\bibinfo  {journal} {Flow, Turbulence
  and Combustion}\ }\textbf {\bibinfo {volume} {72}},\ \bibinfo {pages}
  {115--131} (\bibinfo {year} {2004})}\BibitemShut {NoStop}%
\bibitem [{\citenamefont {Tatarskii}(1971)}]{Tatarskii1971}%
  \BibitemOpen
  \bibfield  {author} {\bibinfo {author} {\bibfnamefont {V.I.}\ \bibnamefont
  {Tatarskii}},\ }\href@noop {} {\emph {\bibinfo {title} {{The effects of the
  turbulent atmosphere on wave propagation}}}}\ (\bibinfo  {publisher} {Israel
  Program for Scientific Translations},\ \bibinfo {address} {Jerusalem},\
  \bibinfo {year} {1971})\BibitemShut {NoStop}%
\bibitem [{\citenamefont {Wilson}\ \emph {et~al.}(1996)\citenamefont {Wilson},
  \citenamefont {Wyngaard},\ and\ \citenamefont {Havelock}}]{Wilson1996}%
  \BibitemOpen
  \bibfield  {author} {\bibinfo {author} {\bibfnamefont {D~Keith}\ \bibnamefont
  {Wilson}}, \bibinfo {author} {\bibfnamefont {John~C}\ \bibnamefont
  {Wyngaard}}, \ and\ \bibinfo {author} {\bibfnamefont {David~I}\ \bibnamefont
  {Havelock}},\ }\bibfield  {title} {\enquote {\bibinfo {title} {{The effect of
  turbulent intermittency on scattering into an acoustic shadow zone}},}\
  }\href@noop {} {\bibfield  {journal} {\bibinfo  {journal} {The Journal of the
  Acoustical Society of America}\ }\textbf {\bibinfo {volume} {99}},\ \bibinfo
  {pages} {3393--3400} (\bibinfo {year} {1996})}\BibitemShut {NoStop}%
\bibitem [{\citenamefont {Kolmogorov}(1941)}]{Kolmogorov1941a}%
  \BibitemOpen
  \bibfield  {author} {\bibinfo {author} {\bibfnamefont {A.~N.}\ \bibnamefont
  {Kolmogorov}},\ }\bibfield  {title} {\enquote {\bibinfo {title} {{The Local
  Structure of Turbulence in Incompressible Viscous Fluid for Very Large
  Reynolds Numbers}},}\ }\href@noop {} {\bibfield  {journal} {\bibinfo
  {journal} {Dokl. Akad. Nauk SSSR}\ }\textbf {\bibinfo {volume} {30}},\
  \bibinfo {pages} {299--303} (\bibinfo {year} {1941})}\BibitemShut {NoStop}%
\bibitem [{\citenamefont {Anselmet}\ \emph {et~al.}(1984)\citenamefont
  {Anselmet}, \citenamefont {Gagne}, \citenamefont {Hopfinger},\ and\
  \citenamefont {Antonia}}]{Anselmet1984}%
  \BibitemOpen
  \bibfield  {author} {\bibinfo {author} {\bibfnamefont {F.}~\bibnamefont
  {Anselmet}}, \bibinfo {author} {\bibfnamefont {Y}~\bibnamefont {Gagne}},
  \bibinfo {author} {\bibfnamefont {E~J}\ \bibnamefont {Hopfinger}}, \ and\
  \bibinfo {author} {\bibfnamefont {R.~A.}\ \bibnamefont {Antonia}},\
  }\bibfield  {title} {\enquote {\bibinfo {title} {{High-order velocity
  structure functions in turbulent shear flows}},}\ }\href@noop {} {\bibfield
  {journal} {\bibinfo  {journal} {Journal of Fluid Mechanics}\ }\textbf
  {\bibinfo {volume} {140}},\ \bibinfo {pages} {63} (\bibinfo {year}
  {1984})}\BibitemShut {NoStop}%
\bibitem [{\citenamefont {Nelkin}(1994)}]{Nelkin1994}%
  \BibitemOpen
  \bibfield  {author} {\bibinfo {author} {\bibfnamefont {Mark}\ \bibnamefont
  {Nelkin}},\ }\bibfield  {title} {\enquote {\bibinfo {title} {{Universality
  and scaling in fully developed turbulence}},}\ }\href@noop {} {\bibfield
  {journal} {\bibinfo  {journal} {Advances in Physics}\ }\textbf {\bibinfo
  {volume} {43}},\ \bibinfo {pages} {143--181} (\bibinfo {year}
  {1994})}\BibitemShut {NoStop}%
\bibitem [{\citenamefont {Landau}(1987)}]{Landau1987}%
  \BibitemOpen
  \bibfield  {author} {\bibinfo {author} {\bibfnamefont {L~D}\ \bibnamefont
  {Landau}},\ }\href@noop {} {\emph {\bibinfo {title} {{Fluid mechanics}}}}\
  (\bibinfo  {publisher} {Pergamon Press},\ \bibinfo {address} {Oxford, England
  New York},\ \bibinfo {year} {1987})\ p.\ \bibinfo {pages} {140}\BibitemShut
  {NoStop}%
\bibitem [{\citenamefont {Oboukhov}(1962)}]{Oboukhov1962}%
  \BibitemOpen
  \bibfield  {author} {\bibinfo {author} {\bibfnamefont {A.~M.}\ \bibnamefont
  {Oboukhov}},\ }\bibfield  {title} {\enquote {\bibinfo {title} {{Some specific
  features of atmospheric tubulence}},}\ }\href@noop {} {\bibfield  {journal}
  {\bibinfo  {journal} {Journal of Fluid Mechanics}\ }\textbf {\bibinfo
  {volume} {13}},\ \bibinfo {pages} {77} (\bibinfo {year} {1962})}\BibitemShut
  {NoStop}%
\bibitem [{Note1()}]{Note1}%
  \BibitemOpen
  \bibinfo {note} {This definition is consistent with Oboukhov's formulation;
  in Kolmogorov's formulation, the averaging volume is a sphere of radius $r$
  centred at $\protect \bm {x}$.}\BibitemShut {Stop}%
\bibitem [{\citenamefont {Stolovitzky}\ \emph {et~al.}(1992)\citenamefont
  {Stolovitzky}, \citenamefont {Kailasnath},\ and\ \citenamefont
  {Sreenivasan}}]{Stolovitzky1992}%
  \BibitemOpen
  \bibfield  {author} {\bibinfo {author} {\bibfnamefont {G.}~\bibnamefont
  {Stolovitzky}}, \bibinfo {author} {\bibfnamefont {P.}~\bibnamefont
  {Kailasnath}}, \ and\ \bibinfo {author} {\bibfnamefont {K.~R.}\ \bibnamefont
  {Sreenivasan}},\ }\bibfield  {title} {\enquote {\bibinfo {title}
  {{Kolmogorov's refined similarity hypotheses}},}\ }\href@noop {} {\bibfield
  {journal} {\bibinfo  {journal} {Physical Review Letters}\ }\textbf {\bibinfo
  {volume} {69}},\ \bibinfo {pages} {1178--1181} (\bibinfo {year}
  {1992})}\BibitemShut {NoStop}%
\bibitem [{\citenamefont {Iyer}\ \emph {et~al.}(2017)\citenamefont {Iyer},
  \citenamefont {Sreenivasan},\ and\ \citenamefont {Yeung}}]{Iyer2017}%
  \BibitemOpen
  \bibfield  {author} {\bibinfo {author} {\bibfnamefont {Kartik~P}\
  \bibnamefont {Iyer}}, \bibinfo {author} {\bibfnamefont {Katepalli~R}\
  \bibnamefont {Sreenivasan}}, \ and\ \bibinfo {author} {\bibfnamefont {P~K}\
  \bibnamefont {Yeung}},\ }\bibfield  {title} {\enquote {\bibinfo {title}
  {{Reynolds number scaling of velocity increments in isotropic turbulence}},}\
  }\href@noop {} {\bibfield  {journal} {\bibinfo  {journal} {Physical Review
  E}\ }\textbf {\bibinfo {volume} {95}},\ \bibinfo {pages} {021101} (\bibinfo
  {year} {2017})}\BibitemShut {NoStop}%
\bibitem [{\citenamefont {Thoroddsen}\ and\ \citenamefont {{Van
  Atta}}(1992)}]{Thoroddsen1992}%
  \BibitemOpen
  \bibfield  {author} {\bibinfo {author} {\bibfnamefont {S~T}\ \bibnamefont
  {Thoroddsen}}\ and\ \bibinfo {author} {\bibfnamefont {C.~W.}\ \bibnamefont
  {{Van Atta}}},\ }\bibfield  {title} {\enquote {\bibinfo {title}
  {{Experimental evidence supporting Kolmogorov's refined similarity
  hypothesis}},}\ }\href@noop {} {\bibfield  {journal} {\bibinfo  {journal}
  {Physics of Fluids A: Fluid Dynamics}\ }\textbf {\bibinfo {volume} {4}},\
  \bibinfo {pages} {2592--2594} (\bibinfo {year} {1992})}\BibitemShut {NoStop}%
\bibitem [{\citenamefont {Praskovsky}(1992)}]{Praskovsky1992}%
  \BibitemOpen
  \bibfield  {author} {\bibinfo {author} {\bibfnamefont {Alexander~A}\
  \bibnamefont {Praskovsky}},\ }\bibfield  {title} {\enquote {\bibinfo {title}
  {{Experimental verification of the Kolmogorov refined similarity
  hypothesis}},}\ }\href@noop {} {\bibfield  {journal} {\bibinfo  {journal}
  {Physics of Fluids A: Fluid Dynamics}\ }\textbf {\bibinfo {volume} {4}},\
  \bibinfo {pages} {2589--2591} (\bibinfo {year} {1992})}\BibitemShut {NoStop}%
\bibitem [{\citenamefont {Thoroddsen}(1995)}]{Thoroddsen1995}%
  \BibitemOpen
  \bibfield  {author} {\bibinfo {author} {\bibfnamefont {S.~T.}\ \bibnamefont
  {Thoroddsen}},\ }\bibfield  {title} {\enquote {\bibinfo {title}
  {{Reevaluation of the experimental support for the Kolmogorov refined
  similarity hypothesis}},}\ }\href@noop {} {\bibfield  {journal} {\bibinfo
  {journal} {Physics of Fluids}\ }\textbf {\bibinfo {volume} {7}},\ \bibinfo
  {pages} {691--693} (\bibinfo {year} {1995})}\BibitemShut {NoStop}%
\bibitem [{\citenamefont {Qian}(1996)}]{Qian1996}%
  \BibitemOpen
  \bibfield  {author} {\bibinfo {author} {\bibfnamefont {J.}~\bibnamefont
  {Qian}},\ }\bibfield  {title} {\enquote {\bibinfo {title} {{Correlation
  coefficients between the velocity difference and local average dissipation of
  turbulence}},}\ }\href@noop {} {\bibfield  {journal} {\bibinfo  {journal}
  {Physical Review E - Statistical Physics, Plasmas, Fluids, and Related
  Interdisciplinary Topics}\ }\textbf {\bibinfo {volume} {54}},\ \bibinfo
  {pages} {981--984} (\bibinfo {year} {1996})}\BibitemShut {NoStop}%
\bibitem [{\citenamefont {Wang}\ \emph {et~al.}(1996)\citenamefont {Wang},
  \citenamefont {Chen}, \citenamefont {Brasseur},\ and\ \citenamefont
  {Wyngaard}}]{Wang1996}%
  \BibitemOpen
  \bibfield  {author} {\bibinfo {author} {\bibfnamefont {Lian-Ping}\
  \bibnamefont {Wang}}, \bibinfo {author} {\bibfnamefont {Shiyi}\ \bibnamefont
  {Chen}}, \bibinfo {author} {\bibfnamefont {James~G.}\ \bibnamefont
  {Brasseur}}, \ and\ \bibinfo {author} {\bibfnamefont {John~C.}\ \bibnamefont
  {Wyngaard}},\ }\bibfield  {title} {\enquote {\bibinfo {title} {{Examination
  of hypotheses in the Kolmogorov refined turbulence theory through
  high-resolution simulations. Part 1. Velocity field}},}\ }\href@noop {}
  {\bibfield  {journal} {\bibinfo  {journal} {Journal of Fluid Mechanics}\
  }\textbf {\bibinfo {volume} {309}},\ \bibinfo {pages} {113} (\bibinfo {year}
  {1996})}\BibitemShut {NoStop}%
\bibitem [{\citenamefont {Chen}\ \emph {et~al.}(1995)\citenamefont {Chen},
  \citenamefont {Doolen}, \citenamefont {Kraichnan},\ and\ \citenamefont
  {Wang}}]{Chen1995}%
  \BibitemOpen
  \bibfield  {author} {\bibinfo {author} {\bibfnamefont {Shiyi}\ \bibnamefont
  {Chen}}, \bibinfo {author} {\bibfnamefont {Gary~D.}\ \bibnamefont {Doolen}},
  \bibinfo {author} {\bibfnamefont {Robert~H.}\ \bibnamefont {Kraichnan}}, \
  and\ \bibinfo {author} {\bibfnamefont {Lian-Ping}\ \bibnamefont {Wang}},\
  }\bibfield  {title} {\enquote {\bibinfo {title} {{Is the Kolmogorov Refined
  Similarity Relation Dynamic or Kinematic?}}}\ }\href@noop {} {\bibfield
  {journal} {\bibinfo  {journal} {Physical Review Letters}\ }\textbf {\bibinfo
  {volume} {74}},\ \bibinfo {pages} {1755--1758} (\bibinfo {year}
  {1995})}\BibitemShut {NoStop}%
\bibitem [{\citenamefont {Chen}\ \emph {et~al.}(1993)\citenamefont {Chen},
  \citenamefont {Doolen}, \citenamefont {Kraichnan},\ and\ \citenamefont
  {She}}]{Chen1993}%
  \BibitemOpen
  \bibfield  {author} {\bibinfo {author} {\bibfnamefont {Shiyi}\ \bibnamefont
  {Chen}}, \bibinfo {author} {\bibfnamefont {Gary~D}\ \bibnamefont {Doolen}},
  \bibinfo {author} {\bibfnamefont {Robert~H}\ \bibnamefont {Kraichnan}}, \
  and\ \bibinfo {author} {\bibfnamefont {Zhen-su}\ \bibnamefont {She}},\
  }\bibfield  {title} {\enquote {\bibinfo {title} {{On statistical correlations
  between velocity increments and locally averaged dissipation in homogeneous
  turbulence}},}\ }\href@noop {} {\bibfield  {journal} {\bibinfo  {journal}
  {Physics of Fluids A: Fluid Dynamics}\ }\textbf {\bibinfo {volume} {5}},\
  \bibinfo {pages} {458--463} (\bibinfo {year} {1993})}\BibitemShut {NoStop}%
\bibitem [{\citenamefont {Schumacher}\ \emph {et~al.}(2007)\citenamefont
  {Schumacher}, \citenamefont {Sreenivasan},\ and\ \citenamefont
  {Yakhot}}]{Schumacher2007}%
  \BibitemOpen
  \bibfield  {author} {\bibinfo {author} {\bibfnamefont {J{\"{o}}rg}\
  \bibnamefont {Schumacher}}, \bibinfo {author} {\bibfnamefont {Katepalli~R}\
  \bibnamefont {Sreenivasan}}, \ and\ \bibinfo {author} {\bibfnamefont
  {Victor}\ \bibnamefont {Yakhot}},\ }\bibfield  {title} {\enquote {\bibinfo
  {title} {{Asymptotic exponents from low-Reynolds-number flows}},}\
  }\href@noop {} {\bibfield  {journal} {\bibinfo  {journal} {New Journal of
  Physics}\ }\textbf {\bibinfo {volume} {9}},\ \bibinfo {pages} {89--89}
  (\bibinfo {year} {2007})}\BibitemShut {NoStop}%
\bibitem [{\citenamefont {Iyer}\ \emph {et~al.}(2015)\citenamefont {Iyer},
  \citenamefont {Sreenivasan},\ and\ \citenamefont {Yeung}}]{Iyer2015}%
  \BibitemOpen
  \bibfield  {author} {\bibinfo {author} {\bibfnamefont {K.~P.}\ \bibnamefont
  {Iyer}}, \bibinfo {author} {\bibfnamefont {K.~R.}\ \bibnamefont
  {Sreenivasan}}, \ and\ \bibinfo {author} {\bibfnamefont {P.~K.}\ \bibnamefont
  {Yeung}},\ }\bibfield  {title} {\enquote {\bibinfo {title} {{Refined
  similarity hypothesis using three-dimensional local averages}},}\ }\href@noop
  {} {\bibfield  {journal} {\bibinfo  {journal} {Physical Review E}\ }\textbf
  {\bibinfo {volume} {92}},\ \bibinfo {pages} {063024} (\bibinfo {year}
  {2015})}\BibitemShut {NoStop}%
\bibitem [{\citenamefont {Lawson}\ and\ \citenamefont
  {Dawson}(2014)}]{Lawson2014}%
  \BibitemOpen
  \bibfield  {author} {\bibinfo {author} {\bibfnamefont {John~M.}\ \bibnamefont
  {Lawson}}\ and\ \bibinfo {author} {\bibfnamefont {James~R.}\ \bibnamefont
  {Dawson}},\ }\bibfield  {title} {\enquote {\bibinfo {title} {{A scanning PIV
  method for fine-scale turbulence measurements}},}\ }\href@noop {} {\bibfield
  {journal} {\bibinfo  {journal} {Experiments in Fluids}\ }\textbf {\bibinfo
  {volume} {55}},\ \bibinfo {pages} {1857} (\bibinfo {year}
  {2014})}\BibitemShut {NoStop}%
\bibitem [{\citenamefont {Cardesa}\ \emph {et~al.}(2017)\citenamefont
  {Cardesa}, \citenamefont {Vela-Mart{\'{i}}n},\ and\ \citenamefont
  {Jim{\'{e}}nez}}]{Cardesa2017}%
  \BibitemOpen
  \bibfield  {author} {\bibinfo {author} {\bibfnamefont {Jos{\'{e}}~I.}\
  \bibnamefont {Cardesa}}, \bibinfo {author} {\bibfnamefont {Alberto}\
  \bibnamefont {Vela-Mart{\'{i}}n}}, \ and\ \bibinfo {author} {\bibfnamefont
  {Javier}\ \bibnamefont {Jim{\'{e}}nez}},\ }\bibfield  {title} {\enquote
  {\bibinfo {title} {{The turbulent cascade in five dimensions}},}\ }\href@noop
  {} {\bibfield  {journal} {\bibinfo  {journal} {Science}\ }\textbf {\bibinfo
  {volume} {357}},\ \bibinfo {pages} {782--784} (\bibinfo {year}
  {2017})}\BibitemShut {NoStop}%
\bibitem [{\citenamefont {Xu}\ \emph {et~al.}(2007)\citenamefont {Xu},
  \citenamefont {Ouellette}, \citenamefont {Vincenzi},\ and\ \citenamefont
  {Bodenschatz}}]{Xu2007}%
  \BibitemOpen
  \bibfield  {author} {\bibinfo {author} {\bibfnamefont {Haitao}\ \bibnamefont
  {Xu}}, \bibinfo {author} {\bibfnamefont {Nicholas~T}\ \bibnamefont
  {Ouellette}}, \bibinfo {author} {\bibfnamefont {Dario}\ \bibnamefont
  {Vincenzi}}, \ and\ \bibinfo {author} {\bibfnamefont {Eberhard}\ \bibnamefont
  {Bodenschatz}},\ }\bibfield  {title} {\enquote {\bibinfo {title}
  {{Acceleration Correlations and Pressure Structure Functions in High-Reynolds
  Number Turbulence}},}\ }\href@noop {} {\bibfield  {journal} {\bibinfo
  {journal} {Physical Review Letters}\ }\textbf {\bibinfo {volume} {99}},\
  \bibinfo {pages} {204501} (\bibinfo {year} {2007})}\BibitemShut {NoStop}%
\bibitem [{\citenamefont {Xu}\ \emph {et~al.}(2011)\citenamefont {Xu},
  \citenamefont {Pumir},\ and\ \citenamefont {Bodenschatz}}]{Xu2011}%
  \BibitemOpen
  \bibfield  {author} {\bibinfo {author} {\bibfnamefont {Haitao}\ \bibnamefont
  {Xu}}, \bibinfo {author} {\bibfnamefont {Alain}\ \bibnamefont {Pumir}}, \
  and\ \bibinfo {author} {\bibfnamefont {Eberhard}\ \bibnamefont
  {Bodenschatz}},\ }\bibfield  {title} {\enquote {\bibinfo {title} {{The
  pirouette effect in turbulent flows}},}\ }\href@noop {} {\bibfield  {journal}
  {\bibinfo  {journal} {Nature Physics}\ }\textbf {\bibinfo {volume} {7}},\
  \bibinfo {pages} {709--712} (\bibinfo {year} {2011})}\BibitemShut {NoStop}%
\bibitem [{\citenamefont {Raffel}(2007)}]{Raffel2007}%
  \BibitemOpen
  \bibfield  {author} {\bibinfo {author} {\bibfnamefont {Markus}\ \bibnamefont
  {Raffel}},\ }\href@noop {} {\emph {\bibinfo {title} {{Particle image
  velocimetry a practical guide}}}}\ (\bibinfo  {publisher} {Springer},\
  \bibinfo {address} {Heidelberg New York},\ \bibinfo {year}
  {2007})\BibitemShut {NoStop}%
\bibitem [{\citenamefont {Knutsen}\ \emph {et~al.}(2017)\citenamefont
  {Knutsen}, \citenamefont {Lawson}, \citenamefont {Dawson},\ and\
  \citenamefont {Worth}}]{Knutsen2017}%
  \BibitemOpen
  \bibfield  {author} {\bibinfo {author} {\bibfnamefont {Anna~N.}\ \bibnamefont
  {Knutsen}}, \bibinfo {author} {\bibfnamefont {John~M.}\ \bibnamefont
  {Lawson}}, \bibinfo {author} {\bibfnamefont {James~R.}\ \bibnamefont
  {Dawson}}, \ and\ \bibinfo {author} {\bibfnamefont {Nicholas~A.}\
  \bibnamefont {Worth}},\ }\bibfield  {title} {\enquote {\bibinfo {title} {{A
  laser sheet self-calibration method for scanning PIV}},}\ }\href@noop {}
  {\bibfield  {journal} {\bibinfo  {journal} {Experiments in Fluids}\ }\textbf
  {\bibinfo {volume} {58}},\ \bibinfo {pages} {1--13} (\bibinfo {year}
  {2017})}\BibitemShut {NoStop}%
\bibitem [{\citenamefont {Borue}\ and\ \citenamefont
  {Orszag}(1996)}]{BorueOrszag1996}%
  \BibitemOpen
  \bibfield  {author} {\bibinfo {author} {\bibfnamefont {Vadim}\ \bibnamefont
  {Borue}}\ and\ \bibinfo {author} {\bibfnamefont {Steven~A.}\ \bibnamefont
  {Orszag}},\ }\bibfield  {title} {\enquote {\bibinfo {title} {{Kolmogorov's
  refined similarity hypothesis for hyperviscous turbulence}},}\ }\href@noop {}
  {\bibfield  {journal} {\bibinfo  {journal} {Physical Review E}\ }\textbf
  {\bibinfo {volume} {53}},\ \bibinfo {pages} {R21--R24} (\bibinfo {year}
  {1996})}\BibitemShut {NoStop}%
\bibitem [{\citenamefont {Voth}\ \emph {et~al.}(1998)\citenamefont {Voth},
  \citenamefont {Satyanarayan},\ and\ \citenamefont {Bodenschatz}}]{Voth1998}%
  \BibitemOpen
  \bibfield  {author} {\bibinfo {author} {\bibfnamefont {Greg~A}\ \bibnamefont
  {Voth}}, \bibinfo {author} {\bibfnamefont {K}~\bibnamefont {Satyanarayan}}, \
  and\ \bibinfo {author} {\bibfnamefont {Eberhard}\ \bibnamefont
  {Bodenschatz}},\ }\bibfield  {title} {\enquote {\bibinfo {title} {{Lagrangian
  acceleration measurements at large Reynolds numbers}},}\ }\href@noop {}
  {\bibfield  {journal} {\bibinfo  {journal} {Physics of Fluids}\ }\textbf
  {\bibinfo {volume} {10}},\ \bibinfo {pages} {2268--2280} (\bibinfo {year}
  {1998})}\BibitemShut {NoStop}%
\bibitem [{\citenamefont {Voth}\ \emph {et~al.}(2002)\citenamefont {Voth},
  \citenamefont {{La Porta}}, \citenamefont {Crawford}, \citenamefont
  {Alexander},\ and\ \citenamefont {Bodenschatz}}]{Voth2002}%
  \BibitemOpen
  \bibfield  {author} {\bibinfo {author} {\bibfnamefont {Greg~A.}\ \bibnamefont
  {Voth}}, \bibinfo {author} {\bibfnamefont {A.}~\bibnamefont {{La Porta}}},
  \bibinfo {author} {\bibfnamefont {Alice~M.}\ \bibnamefont {Crawford}},
  \bibinfo {author} {\bibfnamefont {Jim}\ \bibnamefont {Alexander}}, \ and\
  \bibinfo {author} {\bibfnamefont {Eberhard}\ \bibnamefont {Bodenschatz}},\
  }\bibfield  {title} {\enquote {\bibinfo {title} {{Measurement of particle
  accelerations in fully developed turbulence}},}\ }\href@noop {} {\bibfield
  {journal} {\bibinfo  {journal} {Journal of Fluid Mechanics}\ }\textbf
  {\bibinfo {volume} {469}},\ \bibinfo {pages} {121--160} (\bibinfo {year}
  {2002})}\BibitemShut {NoStop}%
\bibitem [{\citenamefont {Lawson}\ and\ \citenamefont
  {Dawson}(2015)}]{Lawson2015}%
  \BibitemOpen
  \bibfield  {author} {\bibinfo {author} {\bibfnamefont {J.~M.}\ \bibnamefont
  {Lawson}}\ and\ \bibinfo {author} {\bibfnamefont {J.~R.}\ \bibnamefont
  {Dawson}},\ }\bibfield  {title} {\enquote {\bibinfo {title} {{On velocity
  gradient dynamics and turbulent structure}},}\ }\href@noop {} {\bibfield
  {journal} {\bibinfo  {journal} {Journal of Fluid Mechanics}\ }\textbf
  {\bibinfo {volume} {780}},\ \bibinfo {pages} {60--98} (\bibinfo {year}
  {2015})}\BibitemShut {NoStop}%
\bibitem [{\citenamefont {Arad}\ \emph {et~al.}(1998)\citenamefont {Arad},
  \citenamefont {Dhruva}, \citenamefont {Kurien}, \citenamefont {L'vov},
  \citenamefont {Procaccia},\ and\ \citenamefont {Sreenivasan}}]{Arad1998}%
  \BibitemOpen
  \bibfield  {author} {\bibinfo {author} {\bibfnamefont {Itai}\ \bibnamefont
  {Arad}}, \bibinfo {author} {\bibfnamefont {Brindesh}\ \bibnamefont {Dhruva}},
  \bibinfo {author} {\bibfnamefont {Susan}\ \bibnamefont {Kurien}}, \bibinfo
  {author} {\bibfnamefont {Victor~S.}\ \bibnamefont {L'vov}}, \bibinfo {author}
  {\bibfnamefont {Itamar}\ \bibnamefont {Procaccia}}, \ and\ \bibinfo {author}
  {\bibfnamefont {K.~R.}\ \bibnamefont {Sreenivasan}},\ }\bibfield  {title}
  {\enquote {\bibinfo {title} {{Extraction of Anisotropic Contributions in
  Turbulent Flows}},}\ }\href@noop {} {\bibfield  {journal} {\bibinfo
  {journal} {Physical Review Letters}\ }\textbf {\bibinfo {volume} {81}},\
  \bibinfo {pages} {5330--5333} (\bibinfo {year} {1998})}\BibitemShut {NoStop}%
\bibitem [{\citenamefont {Biferale}\ and\ \citenamefont
  {Procaccia}(2005)}]{Biferale2005}%
  \BibitemOpen
  \bibfield  {author} {\bibinfo {author} {\bibfnamefont {Luca}\ \bibnamefont
  {Biferale}}\ and\ \bibinfo {author} {\bibfnamefont {Itamar}\ \bibnamefont
  {Procaccia}},\ }\bibfield  {title} {\enquote {\bibinfo {title} {{Anisotropy
  in turbulent flows and in turbulent transport}},}\ }\href@noop {} {\bibfield
  {journal} {\bibinfo  {journal} {Physics Reports}\ }\textbf {\bibinfo {volume}
  {414}},\ \bibinfo {pages} {43--164} (\bibinfo {year} {2005})}\BibitemShut
  {NoStop}%
\bibitem [{\citenamefont {Arad}\ \emph {et~al.}(1999)\citenamefont {Arad},
  \citenamefont {Biferale}, \citenamefont {Mazzitelli},\ and\ \citenamefont
  {Procaccia}}]{Arad1999}%
  \BibitemOpen
  \bibfield  {author} {\bibinfo {author} {\bibfnamefont {Itai}\ \bibnamefont
  {Arad}}, \bibinfo {author} {\bibfnamefont {Luca}\ \bibnamefont {Biferale}},
  \bibinfo {author} {\bibfnamefont {Irene}\ \bibnamefont {Mazzitelli}}, \ and\
  \bibinfo {author} {\bibfnamefont {Itamar}\ \bibnamefont {Procaccia}},\
  }\bibfield  {title} {\enquote {\bibinfo {title} {{Disentangling Scaling
  Properties in Anisotropic and Inhomogeneous Turbulence}},}\ }\href@noop {}
  {\bibfield  {journal} {\bibinfo  {journal} {Physical Review Letters}\
  }\textbf {\bibinfo {volume} {82}},\ \bibinfo {pages} {5040--5043} (\bibinfo
  {year} {1999})}\BibitemShut {NoStop}%
\bibitem [{\citenamefont {Kurien}\ \emph {et~al.}(2000)\citenamefont {Kurien},
  \citenamefont {L'vov}, \citenamefont {Procaccia},\ and\ \citenamefont
  {Sreenivasan}}]{Kurien2000}%
  \BibitemOpen
  \bibfield  {author} {\bibinfo {author} {\bibfnamefont {Susan}\ \bibnamefont
  {Kurien}}, \bibinfo {author} {\bibfnamefont {Victor~S.}\ \bibnamefont
  {L'vov}}, \bibinfo {author} {\bibfnamefont {Itamar}\ \bibnamefont
  {Procaccia}}, \ and\ \bibinfo {author} {\bibfnamefont {K.~R.}\ \bibnamefont
  {Sreenivasan}},\ }\bibfield  {title} {\enquote {\bibinfo {title} {{Scaling
  structure of the velocity statistics in atmospheric boundary layers}},}\
  }\href@noop {} {\bibfield  {journal} {\bibinfo  {journal} {Physical Review
  E}\ }\textbf {\bibinfo {volume} {61}},\ \bibinfo {pages} {407--421} (\bibinfo
  {year} {2000})}\BibitemShut {NoStop}%
\bibitem [{\citenamefont {Donzis}\ \emph {et~al.}(2008)\citenamefont {Donzis},
  \citenamefont {Yeung},\ and\ \citenamefont {Sreenivasan}}]{Donzis2008}%
  \BibitemOpen
  \bibfield  {author} {\bibinfo {author} {\bibfnamefont {D~A}\ \bibnamefont
  {Donzis}}, \bibinfo {author} {\bibfnamefont {P~K}\ \bibnamefont {Yeung}}, \
  and\ \bibinfo {author} {\bibfnamefont {K~R}\ \bibnamefont {Sreenivasan}},\
  }\bibfield  {title} {\enquote {\bibinfo {title} {{Dissipation and enstrophy
  in isotropic turbulence: Resolution effects and scaling in direct numerical
  simulations}},}\ }\href@noop {} {\bibfield  {journal} {\bibinfo  {journal}
  {Physics of Fluids}\ }\textbf {\bibinfo {volume} {20}},\ \bibinfo {pages}
  {045108} (\bibinfo {year} {2008})}\BibitemShut {NoStop}%
\bibitem [{\citenamefont {Yeung}\ \emph {et~al.}(2015)\citenamefont {Yeung},
  \citenamefont {Zhai},\ and\ \citenamefont {Sreenivasan}}]{Yeung2015}%
  \BibitemOpen
  \bibfield  {author} {\bibinfo {author} {\bibfnamefont {P.~K.}\ \bibnamefont
  {Yeung}}, \bibinfo {author} {\bibfnamefont {X.~M.}\ \bibnamefont {Zhai}}, \
  and\ \bibinfo {author} {\bibfnamefont {Katepalli~R.}\ \bibnamefont
  {Sreenivasan}},\ }\bibfield  {title} {\enquote {\bibinfo {title} {{Extreme
  events in computational turbulence}},}\ }\href@noop {} {\bibfield  {journal}
  {\bibinfo  {journal} {Proceedings of the National Academy of Sciences}\
  }\textbf {\bibinfo {volume} {112}},\ \bibinfo {pages} {12633--12638}
  (\bibinfo {year} {2015})}\BibitemShut {NoStop}%
\end{thebibliography}
%

\end{document}